\documentclass[twocolumn,showpacs,preprintnumbers,amsmath,amssymb]{revtex4}

\usepackage{graphicx}% Include figure files
\usepackage{dcolumn}% Align table columns on decimal point
\usepackage{bm}% bold math
\usepackage{amsmath}

\begin{document}
\title{Immunization
of Susceptible-Infected Model on Scale-Free networks}
\author{Wen-Jie Bai}
\author{Tao Zhou}
\email{zhutou@ustc.edu}
\author{Bing-Hong Wang}
\affiliation{%
Department of Modern Physics, University of Science and Technology
of China, Hefei 230026, PR China
}%

\date{\today}

\begin{abstract}
In this paper, we investigate two major immunization strategies,
random immunization and targeted immunization, of the
susceptible-infected (SI) Model on the Barab\'{a}si-Albert (BA)
networks. For the heterogenous structure, the random strategy is
quite ineffective if the vaccinated proportion is quite small,
while the targeted one which prefers to vaccinate the individuals
with the largest degree can sharply depress the epidemic spreading
even only a small amount of population are vaccinated. The
analytical solution is also obtained,which can capture the trend
of velocity change versus the amount of vaccinated population.
\end{abstract}


\maketitle

\section{Introduction}
Epidemic dynamics, one of the attracting problems in both
biological and physical communities, aims at explaining the
dynamical processes of disease spreading, computer virus
prevailing, and so on. The previous investigations based on the
differential equations are always under the assumption of both
homogeneous infectivity and homogeneous connectivity of each
individual \cite{Anderson1992,Class4}. Therefore, if denote the
possible contacts, along which the infection spreads, by edges,
then the previous studies mainly concentrated on the epidemic
dynamics on complete or random networks. However, against the
above assumption, the empirical data of real networks indicate the
universal existence of heterogeneous topologies
\cite{sf1,sf2,sf3}. One intriguing finding is that many real
networks have approximately power-law degree distributions, that
is to say, the probability distribution function of degree,
$P(k)$, approximately obeys the form $P(k) \sim k^{-\gamma}$, with
$2<\gamma < 3$. This kind of distribution implies an unexpected
abundance of vertices with very large degrees, \emph{i.e.}, the
so-called ``hubs" or ``super-spreaders". This series of networks,
named scale-free (SF) networks, attract many researchers to
investigate the corresponding epidemic behaviors (see the review
paper \cite{Zhou2006a} and the references therein).

The most exemplary models in this field, having been extensively
lucubrated, are the susceptible-infected-susceptible (SIS) and
susceptible-infected-removed (SIR) models. The recent works about
SIS \cite{sis1,sis2} and SIR \cite{sir1,sir2} models on SF
networks present us with completely new epidemic propagation
scenarios that a highly heterogeneous structure will lead to the
absence of any epidemic threshold. However, many real epidemic
processes cannot be properly described by the above two models. In
this paper, we especially focus on the onset dynamics of epidemic
outbreaks, whose behavior is governed by the pure prevalence
without the natural recovery or removing. That is to say, a
speeding time-scale is much smaller than the recovery time-scale.
In the real world, when the speed of the disease is so drastic
that the effect of recovery and death can be ignored, it is the
identical process with above one. In addition, in some
broadcasting processes, each node is in the possession of two
discrete state: received or unreceived. Different from the SIS or
SIR model, the ones having received signals will not alter back to
the state unreceived. Hence, it is more proper to utilize the
so-called susceptible-infected (SI) model to describe those
dynamic processes, in which the infected nodes stay infected and
spread the infection to the susceptible neighbors with rate
$\lambda$.

Very recently, Barth\'elemy \emph{et al.}
\cite{Barthelemy2004,Barthelemy2005} studied the SI model in
Barab\'asi-Albert (BA) networks \cite{Barabasi1999}, and found
that this epidemic process has an infinite spreading velocity in
the limit of infinite population. Following a similar process on
\emph{random Apollonian networks} and weighted scale-free
networks, Zhou \emph{et al.} investigated the effects of
clustering \cite{Zhou2005} and weight distribution \cite{Yan2005}
on SI epidemics. By using the theory of branching processes,
V\'azquez obtained a more accurate solution about the time
behavior of SI model \cite{Vazquez2006}. The SI model with
identical infectivity, which leads to a slower spreading in SF
networks than the standard model, has recently been investigated
by Zhou \emph{et al.} \cite{Zhou2006b}. And the geographical
effect on SI model is studied by Xu \emph{et al.}
\cite{Xu2006a,Xu2006b}. Although these previous works are very
helpful to the deeply understanding of SI epidemic, compared with
the extensively studied SIR and SIS models, the SI model has not
been carefully investigated thus far. Especially, the immunization
effect on SI dynamics, which is very important for controlling the
prevalence, has not yet been investigated. In this paper, we focus
on the immunization effect of SI model on SF networks, which can
be considered as a complementary work of the previous studies on
the immunization of SIR and SIS models.

\section{The Model}
In the standard network SI model, each individual is represented
by a node of the network and the edges are the connections between
individuals along which the infection may spread. Each individual
can be in two discrete states, either susceptible or infected. The
infection transmission is defined by the spreading rate $\lambda$
at which each susceptible individual acquires the infection from
an infected neighbor during one time step.

Using the mean-field theory, the reaction rate equations can be
written as \cite{Barthelemy2004,Barthelemy2005}:
\begin{equation}
\frac{di_{k}(t)}{dt}=\lambda \langle k\rangle
(1-i_{k}(t))\Theta_{k}(t) \label{mean-field1}
\end{equation}
where $i_{k}(t)$ denotes the density of infected individuals with
degree $k$, $\langle k\rangle$ the average degree, and
$\Theta_{k}$ the density of the infected neighbors of a $k$-degree
node. Neglecting terms of order $\mathbb{O}(i^2)$, the evolution
behavior, $i(t)=\sum_ki_k(t)P(k)$, can be approximately solved as
\cite{Barthelemy2004,Barthelemy2005} :
\begin{equation}
i(t)\sim e^{ct},\texttt{  with }c\sim \langle k^2 \rangle/ \langle
k \rangle.
\end{equation}
In an SF networks with a degree distribution exponent
$2<\gamma\leq 3$, the second-order moment $\langle k^2\rangle$
will approach to infinite as the increase of network size $N$,
indicating an infinite velocity in the onset of epidemic
spreading.

\begin{figure}
\scalebox{0.4}[0.4]{\includegraphics{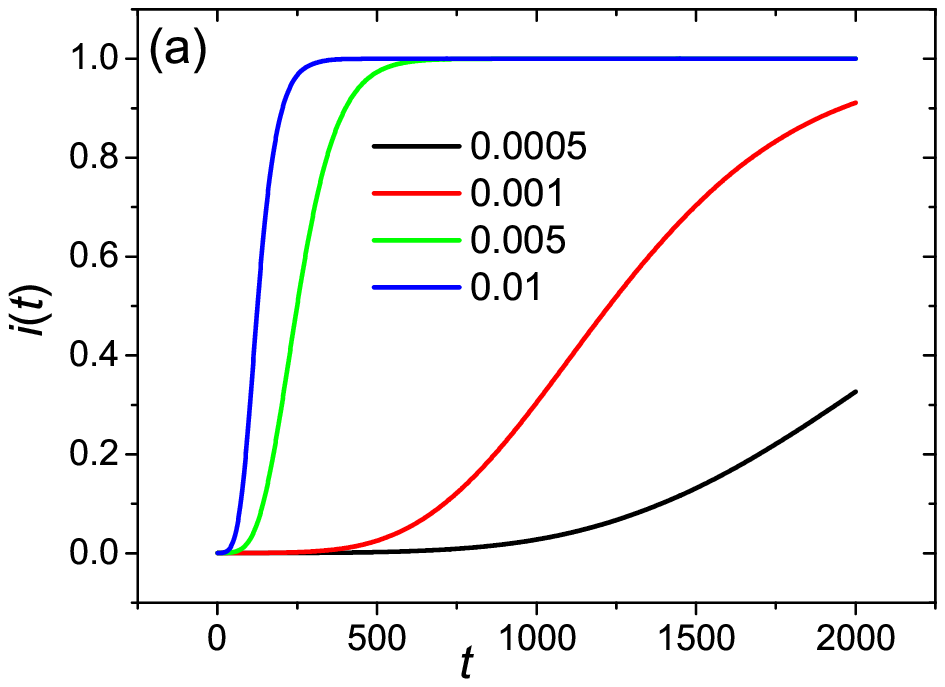}}\scalebox{0.4}[0.4]{\includegraphics{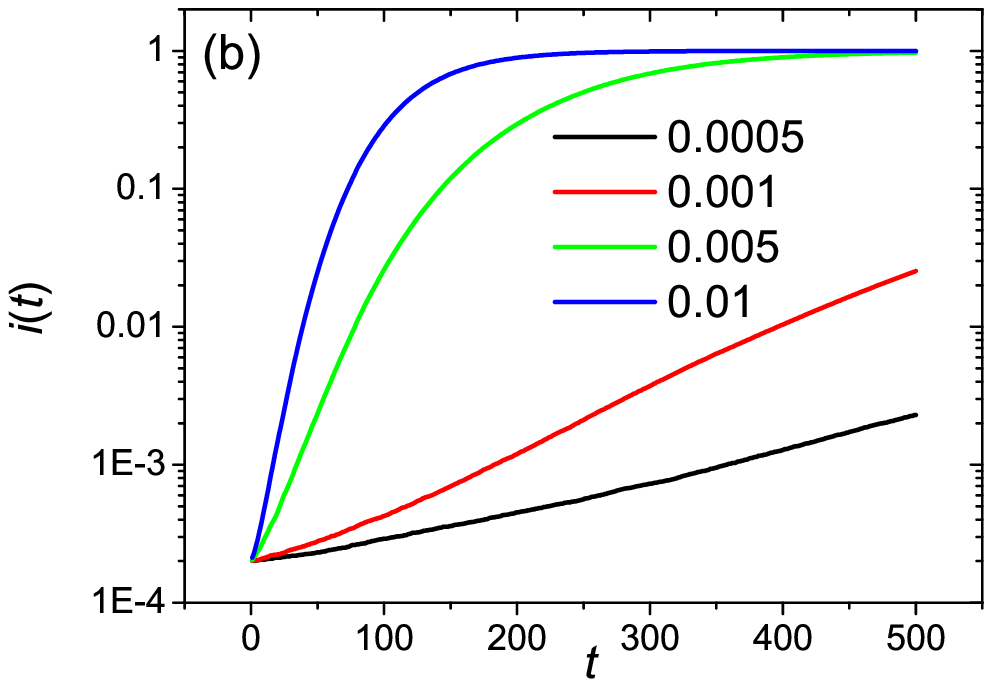}}
\scalebox{0.41}[0.41]{\includegraphics{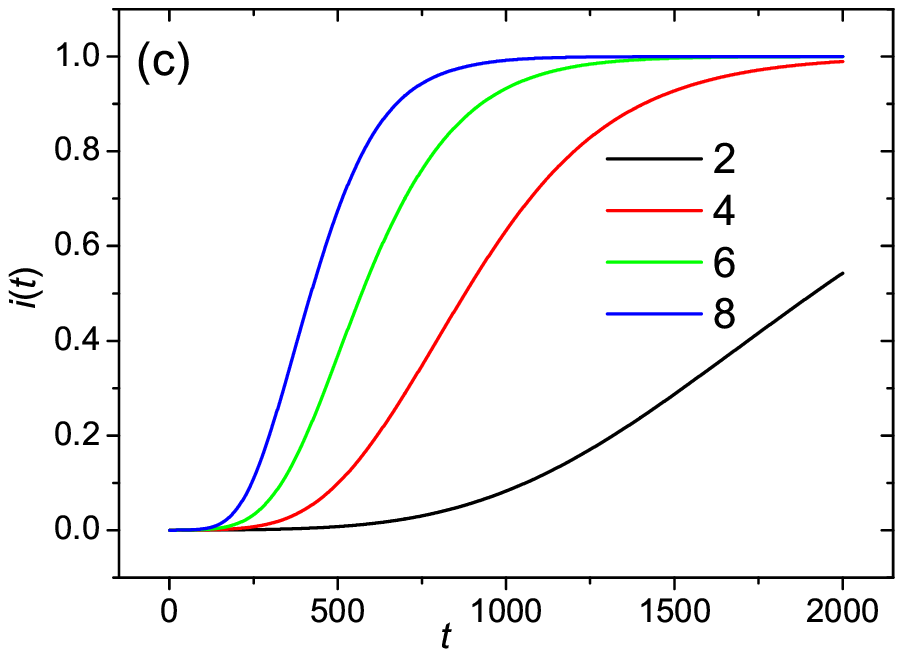}}\scalebox{0.41}[0.41]{\includegraphics{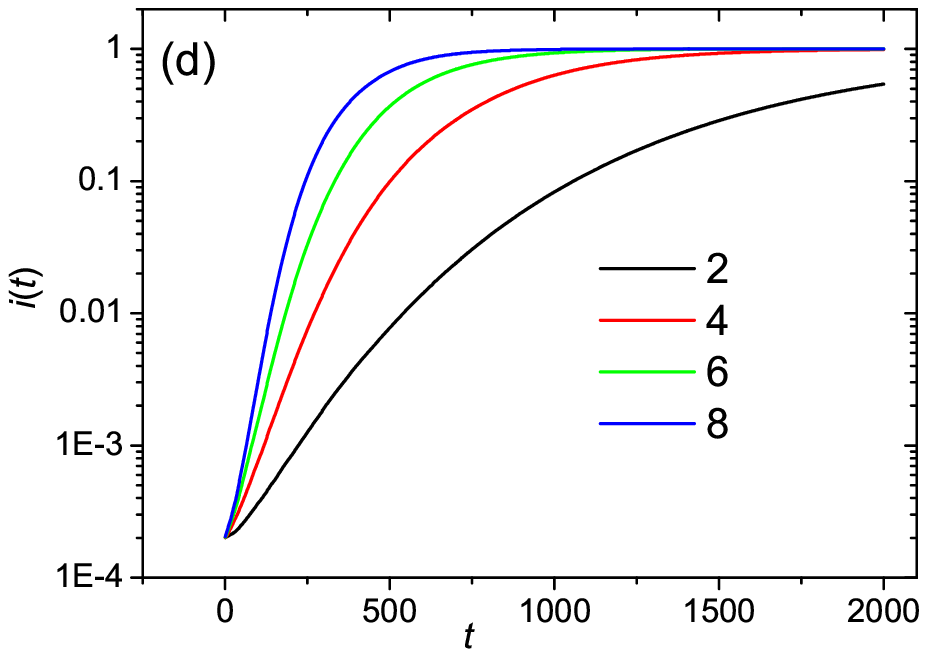}}
\caption{(color online) The infected density $i(t)$ vs time. The
four plots exhibit the time evolution of $i(t)$ for different
spreading rates $\lambda$ in normal (a) and single-log (b)
coordinates, and for different minimal degree $m$ in normal (c)
and single-log (d) coordinates, respectively. The numerical
simulations are implemented based on the BA network of size
$N=5000$. In plots (a) and (b), the average degree of BA networks
is fixed as $\langle k\rangle=6$, and in the plots (c) and (d),
the spreading rate is fixed as $\lambda=0.001$. The legends in
panels (a) and (b) denote the different spreading rates, and the
legends in panels (c) and (d) denote the different minimal
degrees. All the data are averaged over 1000 independent runs.}
\end{figure}

In Fig. 1, we report the simulation results about the time
evolution of infected density with initially one randomly selected
node to be infected. All the simulations are implemented on the BA
networks \cite{Barabasi1999}, which can be constructed by
continuously adding one node with $m$ edges connected to the
existing nodes relying on the probability proportional to their
degrees. The advantage with the BA model is that it is the mostly
studied and lacks structural-biases such as none-zero
degree-degree correlations. Clearly, the epidemic spreading if
very fast, and in the early stage, $i(t)$ follows an exponential
form.

\section{Immunization Effect}
Immunity is a practical controlling strategy to the prevalence of
the disease. The most extensively investigated approaches is the
so-called \emph{mass vaccination} \cite{Anderson1992,ran1} (or
called \emph{random immunization}). In random immunization, a
fraction $f$ of the whole population is randomly selected to be
vaccinated in advance. The most significant problem is that
whether it is effective for the highly heterogeneous networks? In
the previous works, by using the mean-field theory and branch
process theory, Callway \emph{et al.} \cite{Callway2000} and Cohen
\emph{et al.} \cite{Cohen2000}, separately but almost at the same
time, both proved that the random immunization is of less
effectivity for SIR model on SF networks.

\begin{figure}
\scalebox{0.8}[0.8]{\includegraphics{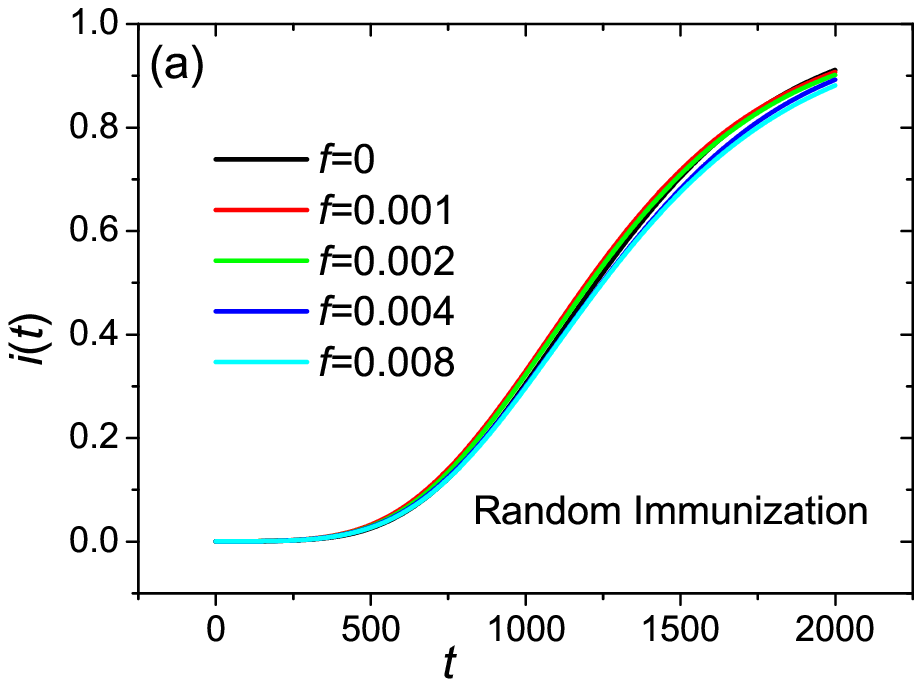}}
\scalebox{0.8}[0.8]{\includegraphics{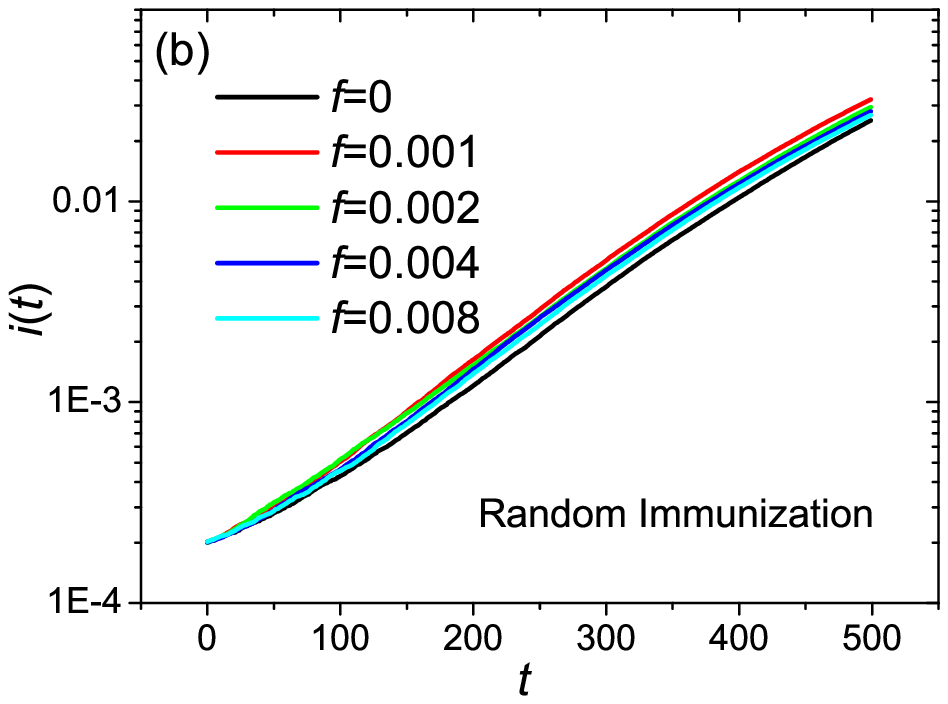}}
\caption{(color online) The infected density $i(t)$ vs time under
random immunization for the SI model based on BA networks in
normal (a) and single-log (b) plots. The network size $N=5000$,
the minimal degree $m=3$, and the spreading rate $\lambda=0.001$
are fixed. The black, red, green, blue and sky-blue curves, from
top to bottom, represent the cases of $f=0$, 0.001, 0.002, 0.004
and 0.008, respectively. All the data are averaged over 1000
independent runs.}
\end{figure}

\begin{figure}
\scalebox{0.8}[0.8]{\includegraphics{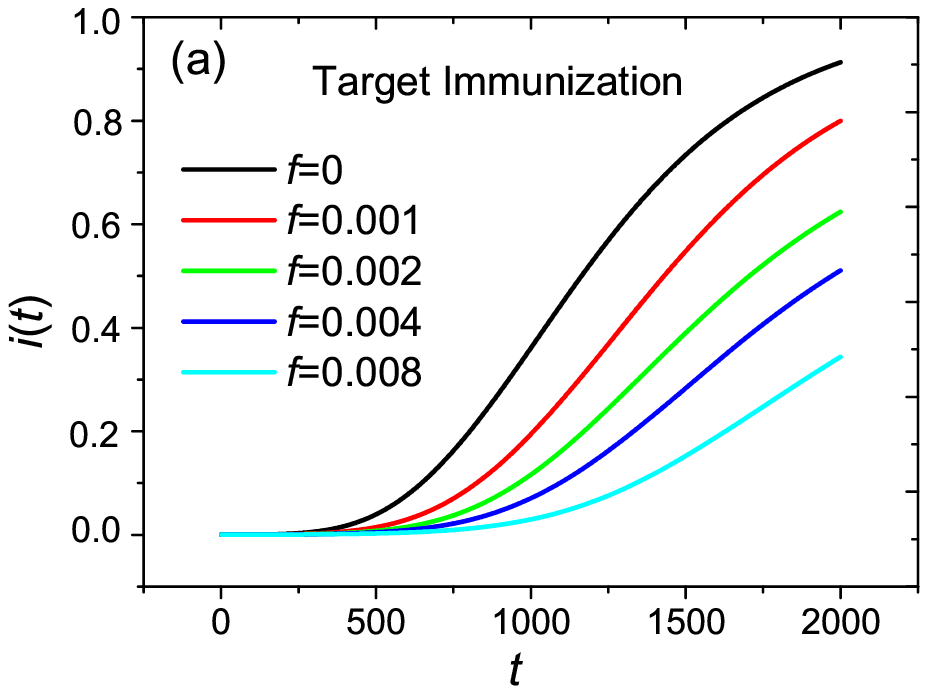}}
\scalebox{0.8}[0.8]{\includegraphics{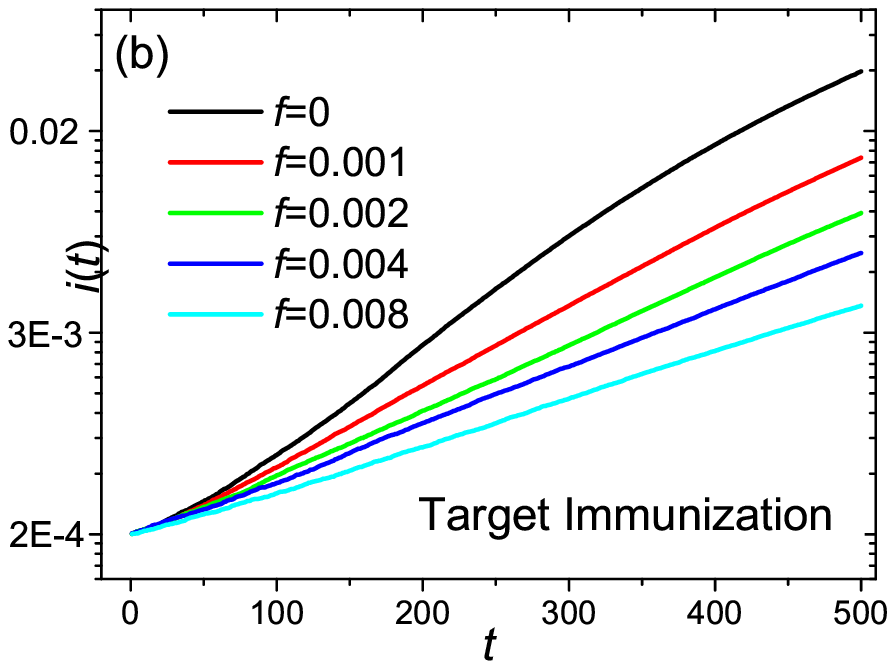}} \caption{ (color
online) The infected density $i(t)$ vs time under targeted
immunization for the SI model based on BA networks in normal (a)
and single-log (b) plots. The network size $N=5000$, the minimal
degree $m=3$, and the spreading rate $\lambda=0.001$ are fixed.
The black, red, green, blue and sky-blue curves, from top to
bottom, represent the cases of $f=0$, 0.001, 0.002, 0.004 and
0.008, respectively. All the data are averaged over 1000
independent runs.}
\end{figure}

In Fig. 2, we plot the time evolution of infected density $i(t)$
for different immunization range $f$, which is defined as the
fraction of population being selected to be vaccinated. From Fig.
2, one can find that the spreading velocity has almost no change
if only a very few individuals are selected to be vaccinated.
Therefore, similar to the situations for SIR model, the random
immunization is of less effectivity for SI model on SF networks.

Other than the random immunization, if the degree of each node is
known, one recently proposed efficient immunization strategy is
the so-called \emph{targeted immunization}
\cite{Pastor2002,Madar2004}, which means to vaccinate the nodes
with the largest degrees first. Fig. 3 shows the effect of
targeted immunization for different $f$. The spreading velocity
remarkably decreases even only a small fraction, $f=0.001$, of
population get vaccinated, which strongly indicates the efficiency
of the targeted immunization. From Fig. 3(b), it is observed that
the time scale governing the epidemic behavior in the early stage
sharply changes even only $10^{-3}$ fraction of population (i.e.
five nodes) are vaccinated.

\begin{figure}
\scalebox{0.8}[0.8]{\includegraphics{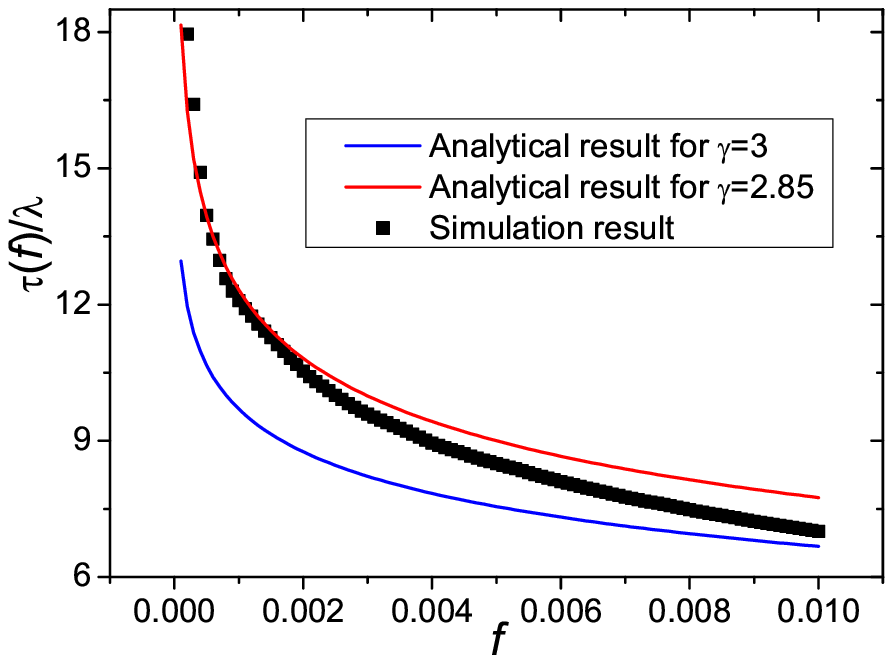}} \caption{(color
online) The rescaled time scale $\tau/\lambda$ vs. immunization
fraction $f$. The black squares represent the simulation result
based on BA networks. The network size $N=10000$, the minimal
degree $m=3$, and the spreading rate $\lambda=0.001$ are fixed.
The blue and red curves denote the analytical results for
$\gamma=3$ and $\gamma=2.85$, respectively. All the data are
averaged over 1000 independent runs.}
\end{figure}

Consider a scale-free network of size $N$, the degree
distribution, $P(k)=Ak^{-\gamma}$, obeys the following normalized
condition
\begin{equation}
\int^{M}_{m}Ak^{-\gamma}dk=1,
\end{equation}
where $A$ is a normalized constant, $M$ the maximal degree and $m$
the minimal degree. We assume after the fraction $f$ of population
with largest degrees having been vaccinated, the maximal degree
decreases to $k_c(f)$, and if $f$ is sufficiently small so that
the degree distribution still obeys a power-law form with exponent
$\gamma$ almost unchanged, then
\begin{equation}
\int^{k_{c}(f)}_{m}Ak^{-\gamma}dk=1-f.
\end{equation}
Following the mean-field theory
\cite{Barthelemy2004,Barthelemy2005}, the time evolution of $i(t)$
in the early stage approximately obeys an exponential form
$i(t)=i(0)e^{\tau(f)t}$, where $i(0)=1/N$ is the initial infected
density. The time scale $\tau(f)$ can be obtained as
\begin{equation}
\tau(f)=\lambda \left(
\frac{\int^{k_{c}(f)}_{m}k^2P'(k)dk}{\int^{k_{c}(f)}_{m}kP'(k)dk}-1
\right),
\end{equation}
where $P'(k)$ is the degree distribution of the network after
vaccination (i.e. after the removal of $Nf$ nodes of largest
degrees), which reads
\begin{equation}
P'(k)=\frac{1}{1-f}P(k),\texttt{    }k=m,m+1,\cdots,k_c(f).
\end{equation}
In the large-limit of $N$, the maximal degree in the original
network, $M\sim \sqrt{N}$, approaches to infinite. Combine Eqs.
(2), (3), (4) and (5), the time-scale $\tau(f)$ after targeted
immunization in the $N\rightarrow \infty$ limit can be
analytically obtained, for any $\gamma \in (2,3)$, as
\begin{equation}
\tau(f)=\lambda \left( m \times \frac{2-\gamma}{3-\gamma} \times
\frac{f^{(3-\gamma)/(1-\gamma)}-1}{f^{(2-\gamma)/(1-\gamma)}-1} -1
\right).
\end{equation}
Especially for the BA networks with $\gamma=3$, the analytical
result is
\begin{equation}
\tau(f)=\lambda \left( \frac{mlnf}{2(\sqrt{f}-1)}-1 \right).
\end{equation}

In Fig. 4, we report the numerical and analytical results for BA
networks with $N=10000$ and $m=3$.  Although the analytical result
for BA networks (e.g. Eq. (8), shown as the blue curve) can
capture the trend of $\tau(f)$, the quantitive departure is very
obvious. Note that, the fitting value of $\gamma$ in finite size
BA networks is smaller than 3.0, which will lead to a even broader
distribution than $P(k)\sim k^{-3}$ thus a faster spreading than
the theoretical prediction. We have obtained the average fitting
value of $\gamma$ as $\overline{\gamma}\approx 2.85$, over 100
independent configurations of BA networks with $N=10000$ and
$m=3$. The red curve in Fig. 4 represent the analytical result for
the modified exponent 2.85 following Eq. (7), one can see clearly
that it agrees well with the simulation for small $f$ and can
capture the trend of $\tau(f)$. For larger $f$, the assumption
that the degree distribution still obeys a power-law form with
same exponent after the removal of $Nf$ hub nodes will not be
valid, resulting in the observed departure.

\section{Conclusion}
As an important branch of the studies on epidemic spreading,
immunity never loses its attraction. Some striking conclusion
somewhat changes our opinions about epidemic. However, despite of
the well-studied SIS and SIR model, the immunization effect on the
outbreaks of epidemic spreading, of significantly practical value,
has not been carefully investigated thus far. The purpose of this
paper is to provide a complementary work of the previous studies
on the immunization of SIR and SIS models.

Two major immunity strategies are investigated based on the BA
networks. The random immunization is of less effectivity while the
targeted immunization can sharply depress the spreading velocity
even only a very few hun nodes are vaccinated. Furthermore, the
analytical results is obtained which agree with the simulation
well for sufficiently small immunization fraction.

\begin{acknowledgments}
This work is funded by the National Basic Research Program of
China (973 Program No.2006CB705500), the National Natural Science
Foundation of China (Grant Nos. 10635040, 70471033, and 10472116),
by the Special Research Funds for Theoretical Physics Frontier
Problems (No. A0524701), and the President Funding of Chinese
Academy of Science.
\end{acknowledgments}

\end{document}